\documentclass[a4paper,11pt]{article}
\pdfoutput=1 

\usepackage{jheppub} 

\usepackage[T1]{fontenc} 
\usepackage{slashed}

\usepackage{cleveref}
\usepackage[numbers,sort&compress]{natbib}
\usepackage{mathtools}
\usepackage{xspace}
\usepackage{xcolor}
\usepackage{multicol}
\usepackage{booktabs}
\usepackage{soul}
\usepackage{makecell}
\usepackage{multirow}
\usepackage{graphicx}
\usepackage{subcaption}
\newcommand{\bea} {\begin{eqnarray}}
\newcommand{\eea} {\end{eqnarray}}
\newcommand{\beq} {\begin{equation}}
\newcommand{\eeq} {\end{equation}}

\allowdisplaybreaks 

\title{\boldmath Probing a light long-lived pseudo-scalar from Higgs decay via displaced taus at the LHC }

\author[a,b]{Lianyou Shan,}
\author[c]{Lei Wang,} 
\author[d,e]{Jin Min Yang,}
\author[d,b]{Rui Zhu}

\affiliation[a]{Institute of High Energy Physics, Chinese Academy of Sciences, Beijing 100049, China}
\affiliation[b]{School of Physical Sciences, University of Chinese Academy of Sciences, Beijing 100049, China}
\affiliation[c]{Department of Physics, Yantai University, Yantai 264005, China}
\affiliation[d]{Institute of Theoretical Physics,  Chinese Academy of Sciences, Beijing 100190, China}
\affiliation[e]{School of Physics, Henan Normal University, Xinxiang 453007, China}

\emailAdd{shanly@ihep.ac.cn}
\emailAdd{leiwang@ytu.edu.cn (corresponding author)}
\emailAdd{jmyang@itp.ac.cn}
\emailAdd{zhurui@itp.ac.cn (corresponding author)}

\abstract{
 A light (GeV mass) long-lived ($c\tau$ around dozens of millimeters) CP-odd scalar can be readily predicted in new physics models. 
In this work we investigate the Higgs decay into such a light scalar plus a $Z$-boson and take the aligned two-Higgs-doublet model (2HDM) as an example. 
This light long-lived scalar, with the dominant decay to tau leptons, will fly over a distance from the production point and present a displaced vertex in an Inner Detector of a generally purposed experiment like ATLAS or CMS. 
In our study we focus on the LHC experiment and perform Monte Carlo simulations for the signal and backgrounds. We demonstrate some benchmark points for the aligned 2HDM and find the signal to be detectable when the luminosity is accumulated to 300 fb$^{-1}$. So our study suggests an experimental search for this process in the ongoing LHC. 
}

\begin{document}
\maketitle
\flushbottom

\section{Introduction}
Numerous measurements of the Higgs properties have been performed at the LHC since its discovery, which basically agree with the Standard Model (SM) predictions. However, some unsolved puzzles, e.g., the cosmic dark matter and baryogenesis, suggest the need of extending the SM Higgs sector (for a recent review, see, e.g., ~\cite{Wang:2023suf}). Among such extensions the two-Higgs-doublet models (2HDMs) (for recent reviews, see, e.g., ~\cite{Wang:2022yhm}) and the low energy supersymmetric models (for recent reviews, see, e.g., ~\cite{Baer:2020kwz,Wang:2022rfd,Yang:2022qyz}) are overwhelmingly popular since they can possibly solve these puzzles\footnote{They can also simultaneously explain the plausible anomalies of muon $g-2$ and the $W$-boson mass (see, e.g., \cite{Han:2022juu,Yang:2022gvz}).} and have rich phenomenology at colliders. In addition to the observed 125 GeV Higgs boson $h$, these models predict extra scalars: one CP-even Higgs boson $H$, one neutral pseudo-scalar $A$, and a pair of charged Higgs bosons $H^{\pm}$.  The ATLAS and CMS collaborations at the LHC have performed intensive searches for new particles from their prompt decays, and so far no significant excess has been observed.
Recently there is an increased interest in the low mass range of long-lived particles (LLPs) and many searches have been performed  at the LHC using an outer detector like a muon spectrometer, a middle detector like a calorimeter or an Inner detector like a tracker (see, e.g., \cite{Knapen:2022afb,ATLAS:2023ian,Wang:2024ieo,Salvatore:2015wpm}).

Theoretically, a light long-lived pseudo-scalar can be readily predicted in some specific 2HDMs or in some non-minimal supersymmetric models like the next-to-minimal supersymmetric model ~\cite{Ellwanger:2009dp,Cao:2012fz,Cao:2013gba,Cao:2010na,Adhikary:2024dix}. 
For example, models like 2HDM+$a$ can provide a very light pseudo scalar $a$ as LLP and has been widely studied \cite{Haisch:2023rqs}. Alternatively in a type-I 2HDM, where the couplings of the pseudo-scalar $A$ with fermions are proportional to $1/\tan\beta$,  a very light $A$ can naturally act as a LLP, as studied  for the experiments like FASER and MATHUSLA ~\cite{Kling:2022uzy,Liu:2024azc}.
In this work, we take however an aligned 2HDM as an example to study a long-lived  pseudo-scalar $A$  produced from the Higgs decay $h\to AZ$ at the LHC. 
In this aligned 2HDM  the Yukawa couplings of $A$, different from the type-I 2HDM, are proportional to three different factors: $\kappa_\ell$ for leptons, $\kappa_d$ for down-type quarks, and $\kappa_u$ for up-type quarks, so $A$ can become a LLP when all Yukawa couplings are sufficiently small. 
In this work we investigate the  decay of $A$ to $\tau^+\tau^-$ for its larger yields (relative to other leptons) and better rejection against backgrounds in experiment (compared with decay to quarks). 
 
This long-lived pseudo-scalar can fly over dozens to hundreds of millimeters once produced in the Higgs decay $h\to AZ$ at the LHC. The tau leptons from its decay ($A\to \tau^+\tau^-$) will further decay into tracks which can be reconstructed only if they traverse sufficient layers in an Inner Detector (Tracker). This has been implemented in the ATLAS and CMS experiments as a special technique of tracking~\cite{ATLAS:2023nze,CMS:2014pgm}. Upon these Large Radius Tracks (LRT), it becomes even possible to reconstruct the decaying vertex of the LLP~\cite{ATLAS:2019wqx,CMS:2021tkn}. In this work we will use these new techniques as performed in ~\cite{ATLAS:2020xyo}. On the other hand, the tau jet with a displaced track provides an outstanding signature to our signal. Since the light pseudo-scalar is produced from the decay of a not-so-heavy Higgs boson ($\sim 125$ GeV ) in association with a rather heavy $Z$-boson ($\sim 91$ GeV), its momentum is usually not large, such that its decay products are usually soft. Considering the lower velocity of the taus from the pseudo-scalar decay, in this work we propose to reconstruct a merged (joint) vertex from the decays of the two taus in order to suppress the so-called pileup backgrounds. Although this reconstruction has larger uncertainties than the conventional (standard) vertexing, as will be demonstrated in our study, it can help to discriminate our signal from backgrounds. 

This paper is organized as follows.
In Sec.~\ref{sec:2HDM} we introduce an aligned two-Higgs-doublet model containing a light CP-odd scalar. After 
showing the relevant theoretical and experimental constraints on this model, we demonstrate the allowed parameter space for a light long-lived pseudo-scalar.  
In Sec.~\ref{sec:bckg_VertexLRT} we focus on the LHC and perform detailed Monte Carlo simulations for the signal and backgrounds. 
Then we show the results for some benchmark points in the aligned two-Higgs-doublet model. 
Finally the conclusions are given in Sec.~\ref{conclusion}.
      
\section{The aligned two-Higgs-doublet model and its phenomenology}
\label{sec:2HDM}
A 2HDM extends the SM simply by adding an additional $SU(2)_L$ Higgs-doublet field,
whose general potential is given by 
\begin{eqnarray}
\label{V2HDM} \mathrm{V}_{2HDM} &=& m^2_{11} (\Phi_1^{\dagger} \Phi_1)
+ m^2_{22} (\Phi_2^{\dagger}
\Phi_2) - \left[m^2_{12} \Phi_1^{\dagger} \Phi_2 + \rm h.c.\right]\nonumber \\
&&+ \frac{\lambda_1}{2}  (\Phi_1^{\dagger} \Phi_1)^2 +
\frac{\lambda_2}{2} (\Phi_2^{\dagger} \Phi_2)^2 + \lambda_3
(\Phi_1^{\dagger} \Phi_1)(\Phi_2^{\dagger} \Phi_2) + \lambda_4
(\Phi_1^{\dagger}
\Phi_2)(\Phi_2^{\dagger} \Phi_1) \nonumber \\
&&+ \left[\frac{\lambda_5}{2} (\Phi_1^{\dagger} \Phi_2)^2  
+  \lambda_6 (\Phi_1^{\dagger} \Phi_1) (\Phi_1^{\dagger} \Phi_2) + \lambda_7 (\Phi_2^{\dagger} \Phi_2) (\Phi_1^{\dagger} \Phi_2) +\rm
h.c.\right].
\end{eqnarray}
Since we do not consider the CP-violating effects in this work (for CP-violation in 2HDM, see, e.g., \cite{Gunion:2005ja}), we assume
all the masses and couplings to be real. For simplicity, we choose $\lambda_6=\lambda_7=0$
in our following study.
Here the fields $\Phi_1$ and $\Phi_2$ are both complex Higgs-doublets with hypercharge $Y = 1$:
\begin{equation}
\Phi_1=\left(\begin{array}{c} \phi_1^+ \\
\frac{1}{\sqrt{2}}\,(v_1+\phi_1+ia_1)
\end{array}\right)\,, \ \ \
\Phi_2=\left(\begin{array}{c} \phi_2^+ \\
\frac{1}{\sqrt{2}}\,(v_2+\phi_2+ia_2)
\end{array}\right),
\end{equation}
where $v_1$ and $v_2$ are respectively the vacuum expectation values (VEVs) of $\phi_1$ and $\phi_2$, and related to the electroweak scale $v = 246$ GeV by $v^2 = v^2_1 + v^2_2$. The ratio of the two VEVs is defined by $\tan\beta=v_2 /v_1$.
The relations between the Higgs masses and the couplings are presented in Appendix \ref{parameter-space}. 

\subsection{The aligned 2HDM model }
 After spontaneous electroweak symmetry breaking, the mass eigenstates are obtained by rotating the original fields:
\begin{eqnarray}
\left(\begin{array}{c}H \\ h \end{array}\right) =  \left(\begin{array}{cc}c_\alpha & s_\alpha \\ -s_\alpha & c_\alpha \end{array}\right)  \left(\begin{array}{c} \phi_1 \\ \phi_2 \end{array}\right) , ~~~
\left(\begin{array}{c}G^0 \\ A \end{array}\right) =  \left(\begin{array}{cc}c_\beta & s_\beta \\ -s_\beta & c_\beta \end{array}\right)  \left(\begin{array}{c} a_1 \\ a_2 \end{array}\right) , 
\label{rot-cp-odd}
\end{eqnarray}
with the shorthand notations defined by $c_\alpha\equiv \cos\alpha$, $s_\alpha\equiv \sin\alpha$, $c_\beta \equiv \cos\beta$, and $s_\beta\equiv \sin\beta$.
The fields $G^0$ is the Nambu-Goldstone boson which is absorbed by gauge boson $Z$ as the longitudinal component. The remaining physical states are two neutral
CP-even states ($h$, $H$), one neutral pseudo-scalar $A$. \footnote{The pair of charged Goldstone and charged Higgs $( G^{\pm}, ~ H^{\pm})$ shares the same rotation as $(G^0, A)$.}  As shown in  Appendix \ref{parameter-space}, it is more convenient to take the physics masses and mixing angles as the independent parameters for phenomenological study:
\begin{equation}
m_h,m_H,m_A,m_{H^\pm},\cos(\beta-\alpha),\tan\beta,m_{12}^2
\end{equation}
The general Yukawa interactions are given by
 \bea\label{gene-Yukawa}
- {\cal L} &=&Y_{u2}\,\overline{Q}_L \, \tilde{{ \Phi}}_2 \,u_R
+\,Y_{d2}\,
\overline{Q}_L\,{\Phi}_2 \, d_R\, + \, Y_{\ell 2}\,\overline{L}_L \, {\Phi}_2\,e_R \,\nonumber\\
&&+Y_{u1}\,\overline{Q}_L \, \tilde{{ \Phi}}_1 \,u_R
+\,Y_{d1}\,
\overline{Q}_L\,{\Phi}_1 \, d_R\, + \, Y_{\ell 1}\,\overline{L}_L \, {\Phi}_1\,e_R+\, \mbox{h.c.}\,,
\eea 
with $Q_L^T=(u_L\,,d_L)$, $L_L^T=(\nu_L\,,l_L)$,
$\widetilde\Phi_{1,2}=i\tau_2 \Phi_{1,2}^*$, and each  $3 \times 3$  $Y$-matrix being in family
space.
The tree-level flavour-changing neutral currents can be eliminated by 
requiring the alignment in flavour space of the Yukawa matrices ~\cite{Pich:2009sp,Wang:2013sha}. 
As detailed in Appendix \ref{aligned}, the two Yukawa matrices coupling to a given type of right-handed fermions are assumed to be proportional to each other, so that the Yukawa coupling for each family (generation), can be described by single parameter $\kappa_f$:
\begin{eqnarray}
- {\cal L}_Y &&= \frac{m_f}{v}~ y_h^f~h\bar{f}f+\frac{m_f}{v}~y_H^f~H\bar{f}f\nonumber\\
&&-i\frac{m_u}{v}\kappa_u  ~A \bar{u} \gamma_5 u + i\frac{m_d}{v}\kappa_d  ~A \bar{d} \gamma_5 d+ i\frac{m_\ell}{v}\kappa_\ell  ~ A \bar{\ell} \gamma_5 \ell\nonumber\\
&&+ H^+~ \bar{u} ~V_{\rm CKM}~ (\frac{\sqrt{2}m_d}{v}\kappa_d P_R  - \frac{\sqrt{2}m_u}{v}\kappa_u P_L)  d + \frac{\sqrt{2}m_\ell}{v}\kappa_\ell H^+~\bar{\nu} P_R  e + h.c.\label{yuka-coupling}
\end{eqnarray}
where $y_h^f=\sin(\beta-\alpha)+\cos(\beta-\alpha)\kappa_f$ and $y_H^f=\cos(\beta-\alpha)-\sin(\beta-\alpha)\kappa_f$ with $f=u,d,\ell$.

The neutral Higgs couplings to $VV$ ($VV\equiv ZZ, WW$) are obtained from the gauge-kinetic Lagrangian
\begin{align}
\mathcal{L}_{SVV} = &\frac{g^2+g'^2}{8}v^2~ZZ~\left(1+2\frac{h}{v}y_h^V+2\frac{H}{v}y_H^V\right) \nonumber\\
                               &+\frac{g^2}{4}v^2~W^+W^-~\left(1+2\frac{h}{v}y_h^V+2\frac{H}{v}y_H^V\right),
\end{align}
where $y_h^V=\sin(\beta-\alpha)$ and $y_H^V=\cos(\beta-\alpha)$.
Because of CP-conservation, the pseudo-scalar $A$ has no couplings to $VV$.
The part of the Lagrangian describing the interaction between one gauge boson and two scalars is
\begin{align}
\mathcal{L}_{SSV} = &\frac{g}{2}W^+_\mu\left((H^-\overset{\leftrightarrow}{\partial}^\mu A)-i\cos(\beta-\alpha)(H^-\overset{\leftrightarrow}{\partial}^\mu h)+i\sin(\beta-\alpha)(H^-\overset{\leftrightarrow}{\partial}^\mu H))+h.c. \right)\notag\\
&+\frac{g}{2c_W}Z_\mu\left(\cos(\beta-\alpha)(A\overset{\leftrightarrow}{\partial}^\mu h)-\sin(\beta-\alpha)(A\overset{\leftrightarrow}{\partial}^\mu H)\right)\notag\\
&+\left(ie\gamma_\mu+i\frac{g(c_W^2-s_W^2)}{2c_W}Z_\mu \right)(H^\mp\overset{\leftrightarrow}{\partial}^\mu H^\pm),
\end{align}
where $c_W = \cos\theta_W$ and $s_W = \sin\theta_W$ with $\theta_W$ denoting the Weinberg angle. Clearly, the coupling of $hAZ$ is proportional to $\cos(\beta - \alpha)$.

\subsection{Current theoretical and experimental constraints}
\label{constraints}
In our discussions, we take the light CP-even scalar $h$ as the observed 125 GeV Higgs boson, and fix the masses of $H$ and $H^\pm$ at 500 GeV. 
We perform a random scan over the following ranges for $m_A$, $\cos(\beta - \alpha)$ and $\tan\beta$:
\begin{equation}
    5{\rm ~GeV} <m_A<30{\rm ~GeV},~0<\cos(\beta-\alpha)<0.4,~0.5<\tan\beta<5.
\label{freeParas}
\end{equation}
From Appendix \ref{tbeta} $\tan\beta$ should be small and here we restrain it within 0.5-5.

In our scan $m_{12}^2$ is also a free parameter whose value is tuned to satisfy the theoretical constraints from the vacuum stability and perturbativity as well as unitarity.   
Because the Yukawa couplings of A are proportional to three
different factors: $\kappa_\ell$ for leptons, $\kappa_d$ for down-type quarks, and $\kappa_u$ for up-type quarks, a light A will decay dominantly to $\tau^+\tau^-$ for $\kappa_\ell \gg \kappa_u$ and $\kappa_\ell \gg \kappa_d$. In addition, $\kappa_u$ and $\kappa_d$ can be stringently constrained by the  searches for additional Higgs bosons via $gg\to H$, $gg \to H\bar{f}f$, $gg\to tbH^-$ and $gb\to tH^-$ production processes at the LHC. Therefore, for the Yukawa sector, we simply set $\kappa_u=\kappa_d$ to zero, and scan $\kappa_\ell$ within a range of small values,
\begin{equation}
   ~10^{-6}<\kappa_\ell<10^{-4}.
\label{onlyTau}
\end{equation}

In addition to the theoretical constraints, we also consider the following experimental constraints:
\begin{itemize}
\item[(i)] {\bf The constraints by the oblique parameters}. The extra Higgs bosons can modify the oblique parameters ($S$, $T$, $U$) from the SM predictions via the loop effects. We employ the \textsc{2HDMC} ~\cite{Eriksson:2009ws} package to calculate the values of $S$, $T$ and $U$, 
and perform a fit to the oblique parameters $S$, $T$, $U$ using the results of  ~\cite{ParticleDataGroup:2020ssz},
\beq
S=-0.01\pm 0.10,~~  T=0.03\pm 0.12,~~ U=0.02 \pm 0.11, 
\eeq
with the correlation coefficients 
\beq
\rho_{ST} = 0.92,~~  \rho_{SU} = -0.80,~~  \rho_{TU} = -0.93.
\eeq


\item[(ii)]   {\bf The constraints by the signal strength of the 125 GeV Higgs.} 
The non-vanishing $\cos(\beta-\alpha)$ in the couplings of $h$ to the SM particles, 
 is constrained by HiggsSignals ~\cite{Bechtle:2013xfa} package to satisfy all the observations (within $2\sigma$). The decay $h\to AA$ is bounded by the invisible Higgs decay of $BR(h\to {\rm invisible}) < 10.7\%$ ~\cite{PhysRevD.110.030001}.


\item[(iii)]  {\bf The constraints from the direct searches for additional Higgs and flavor observables.} 
HiggsBounds~\cite{Bechtle:2008jh} is employed for the exclusion
constraints from the searches for the neutral and charged Higgs bosons at the LEP and the LHC at 95\% confidence level.
Note that our pseudo-scalar is long-lived, so that the existing bounds from the searches of a prompt scalar are not applicable. 
Since we choose $\kappa_d=\kappa_u=0$ and a sufficient small $\kappa_\ell$, and therefore the Yukawa couplings of the extra Higgs bosons ($H$, $H^\pm$, $A$) are very small and even absent according to Eq. (\ref{yuka-coupling}), the scan samples can easily pass the exclusion limits of searches for additional Higgs bosons via $gg\to H$, $gg \to H\bar{f}f$, $gg\to tbH^-$ and $gb\to tH^-$ production processes at the LHC.
Also in this case a pseudo-scalar $A$ with $m_A>$ 3 GeV can survive from flavor measurements like: the $B$-meson decays ~\cite{LHCb:2015nkv,LHCb:2016awg}  and $D$-meson decays ~\cite{LHCb:2020car} from the LHCb,  the Kaon decays from the NA62 ~\cite{NA62:2021zjw} and E949 ~\cite{BNL-E949:2009dza}.


\end{itemize}
\renewcommand{\thesubfigure}{\roman{subfigure}}
\begin{figure}[t]
    \centering
    \begin{subfigure}{0.32\textwidth}
        \centering
        \includegraphics[width=\linewidth]{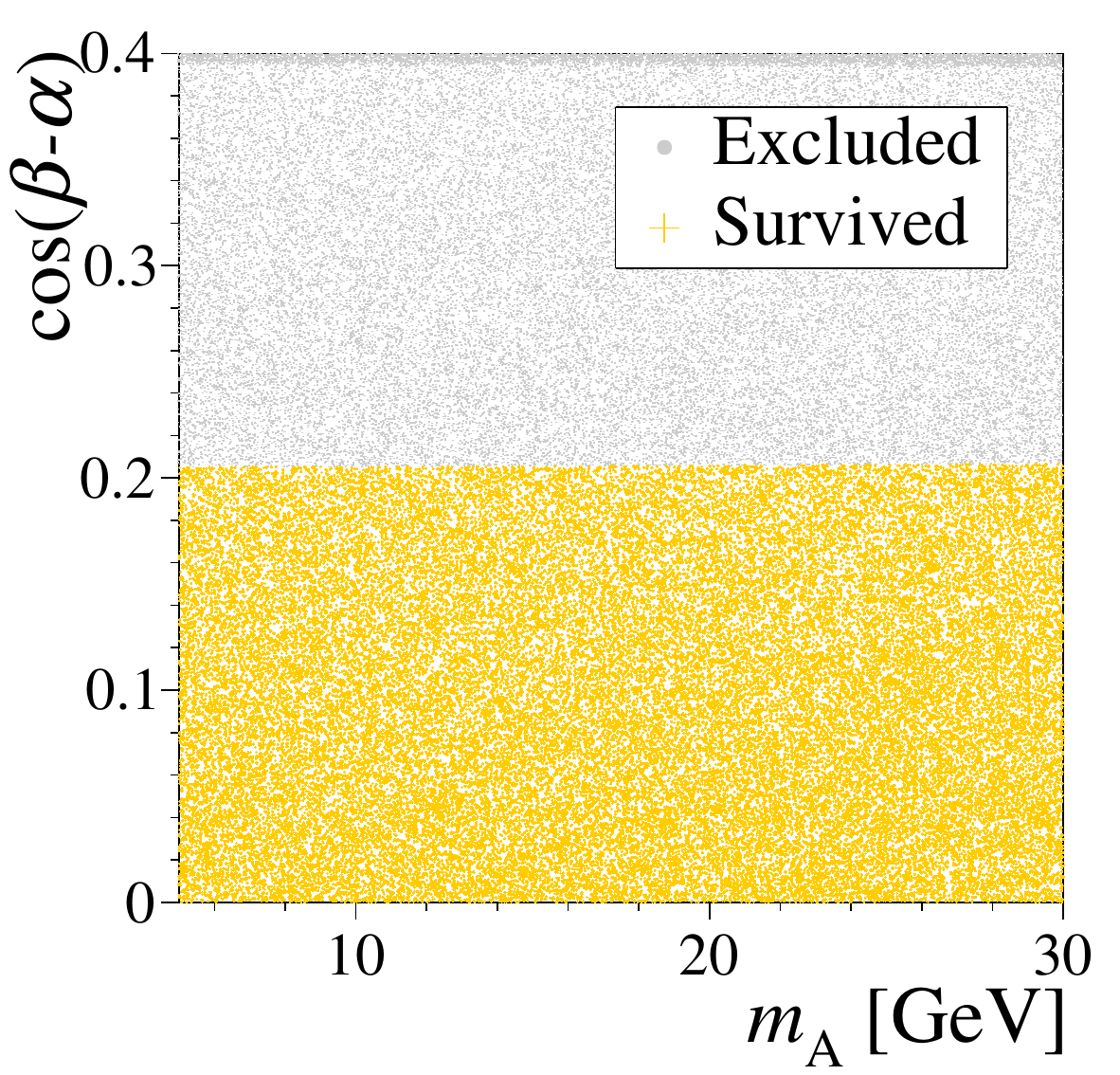}
        \caption{S,T,U}
        \label{fig:sub1}
    \end{subfigure}
    \hfill
    \begin{subfigure}{0.32\textwidth}
        \centering
        \includegraphics[width=\linewidth]{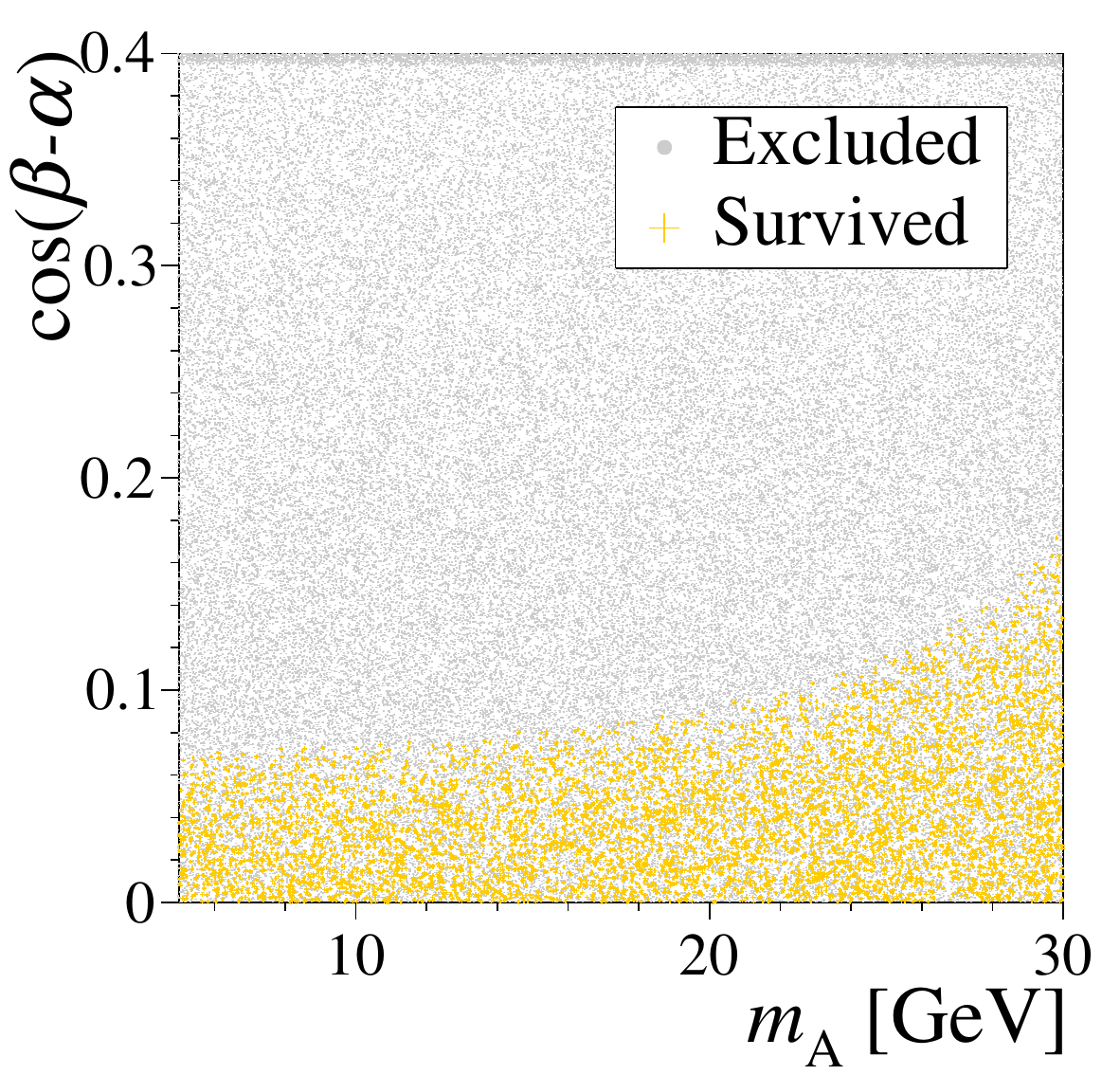}
        \caption{125 GeV Higgs}
        \label{fig:sub2}
    \end{subfigure}
    \hfill
    \begin{subfigure}{0.32\textwidth}
        \centering
        \includegraphics[   width=\linewidth]{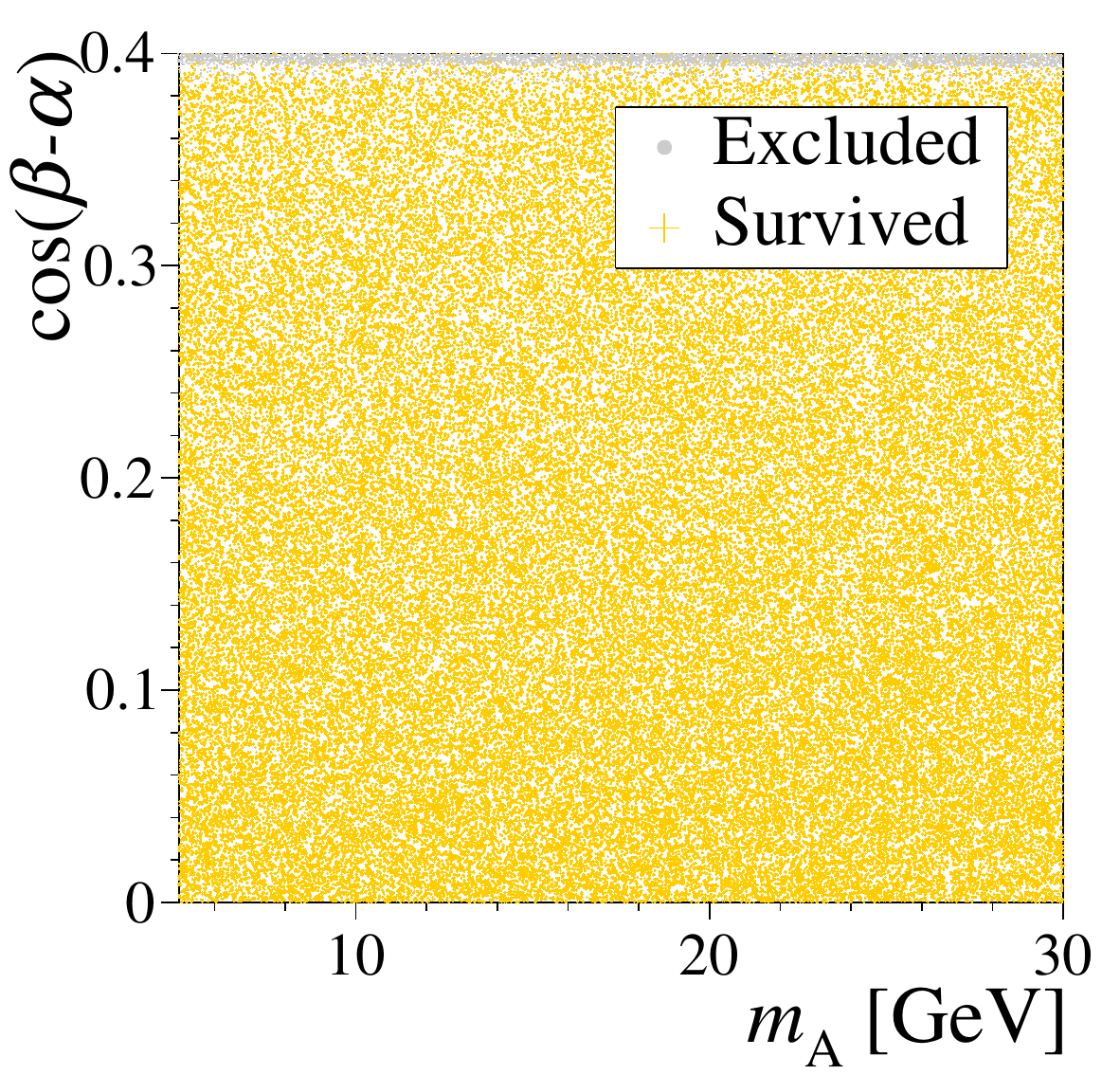}
        \caption{additional Higgs}
        \label{fig:sub3}
    \end{subfigure}
    \caption{The respective exclusion capabilities of experimental constraints (i)-(iii).}
    \label{constraints}
\end{figure}

\renewcommand{\thesubfigure}{\arabic{subfigure}}
\begin{figure}[t]
    \centering
    \begin{subfigure}{0.32\textwidth}
        \centering
        \includegraphics[width=\linewidth]{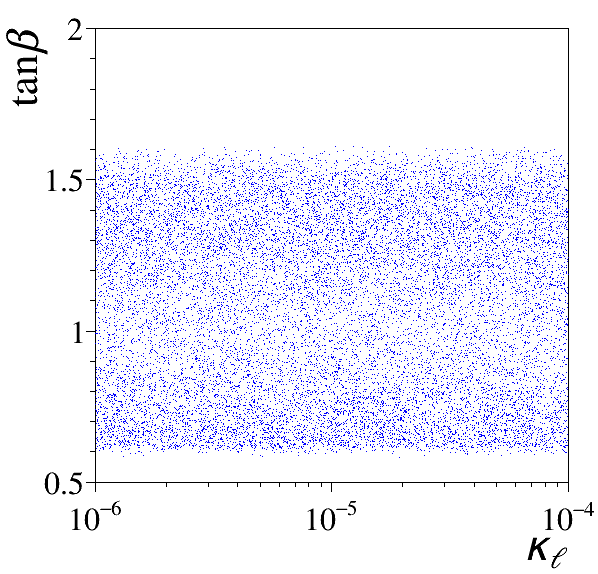}
        \label{fig:sub1}
    \end{subfigure}
    \hfill
    \begin{subfigure}{0.32\textwidth}
        \centering
        \includegraphics[width=\linewidth]{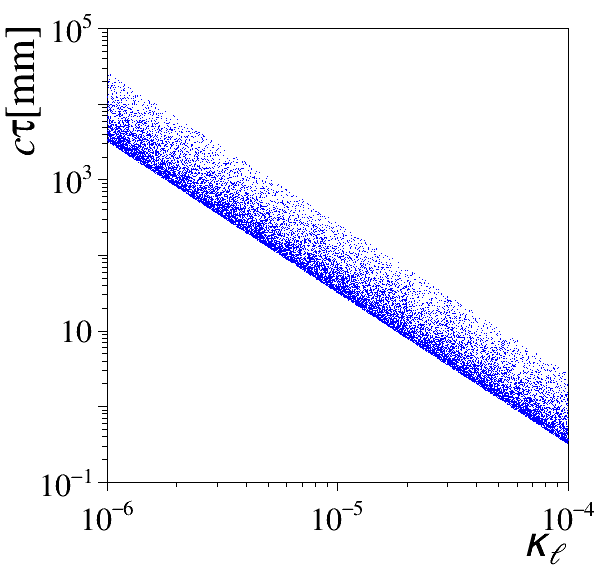}
        \label{fig:sub2}
    \end{subfigure}
    \hfill
    \begin{subfigure}{0.32\textwidth}
        \centering
        \includegraphics[width=\linewidth]{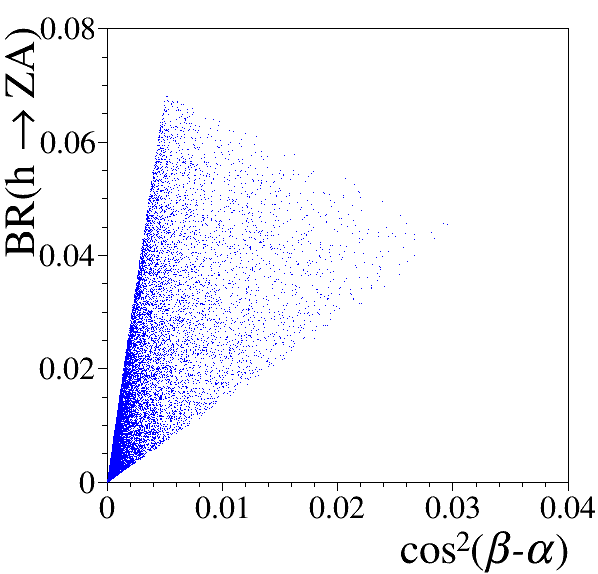}
        \label{fig:sub3}
    \end{subfigure}
    \vspace{-.5cm}
    \caption{The surviving samples of the aligned 2HDM allowed by the constraints (i)-(iii), displayed in different planes of parameters.}
    \label{scan-scatter}
\end{figure}

As shown in Fig.\ref{constraints}, we project the parameter points that satisfy the theoretical constraints onto the cos($\beta-\alpha$)-$m_A$ plane to illustrate the exclusion capability of each experimental constraint. The measurements of the 125 GeV Higgs provide the most stringent constraint, excluding approximately 94$\%$ of the parameter points.

In Fig.\ref{scan-scatter}, we present scatter plots of the samples that satisfy all the aforementioned constraints on several other two-dimensional planes. The following characteristics can be observed from this figure:
\begin{itemize}

\item[(1)] The left panel shows that $\tan\beta$ is favored in the range 0.6 and 1.6, as explained in Appendix \ref{tbeta}.

\item[(2)] The middle panel indicates that the invariant lifetime $c\tau$ of $A$ is primarily determined by $\kappa_\ell$  as expected and is inversely proportional to $\kappa_\ell^2$.
\item[(3)] The right panel shows that the branching ratio of $h$ decay to $A$ and $Z$, i.e $BR(h \to ZA)$, is proportional to $\cos^2(\beta - \alpha)$, with a maximum value of approximately 0.068. Given that the cross-section for Higgs production via gluon fusion equals $48.58$ pb~\cite{LHCHiggsCrossSectionWorkingGroup:2016ypw}, the maximum cross section for $gg \to h \to ZA$ is approximately 3.3 pb.
\end{itemize}

\section{Observability at the LHC }
\label{sec:bckg_VertexLRT}
We focus on the process at the LHC:
\begin{equation}
p p \to h \to Z(\to \ell^+ \ell^-)A(\to \tau^+ \tau^-)
\end{equation}
For this process, the signal has an obvious peak at the $Z$-resonance and at least one hadronic tau with displaced tracks as well as a displaced vertex. With these signatures, the $Z\tau\tau$ production with $\tau\tau$ coming from a virtual $Z$ or a virtual photon  seems to be the dominant background. Whereas, considering the mis-identification of a $b$-jet as a tau jet, the $t\bar{t}$ and $Zbb$ events are also the important backgrounds (actually they are found to surpass the $Z\tau\tau$ background).  

To save computing time in numerical computation, only a few representative benchmark points are chosen from the allowed parameter space for an illustration, as listed in Table \ref{bps}. 
We choose low mass at 5 GeV and high mass at 30 GeV, and vary the invariant lifetime from 3 mm to 300 mm.  
We use \textsc{FeynRules}~\cite{Alloul:2013bka} to generate the UFO model file,  \textsc{MadGraph5\_aMC@NLO}~\cite{Alwall:2014hca, Frederix:2018nkq} to generate parton-level events at the leading order (LO), \textsc{Pythia8}~\cite{Bierlich:2022pfr, Sjostrand:2014zea} for parton showering and hadronization, and then prepare signal events in HepMC3 format.

\begin{figure}[!t]
	\centering
	\includegraphics[width=12cm]{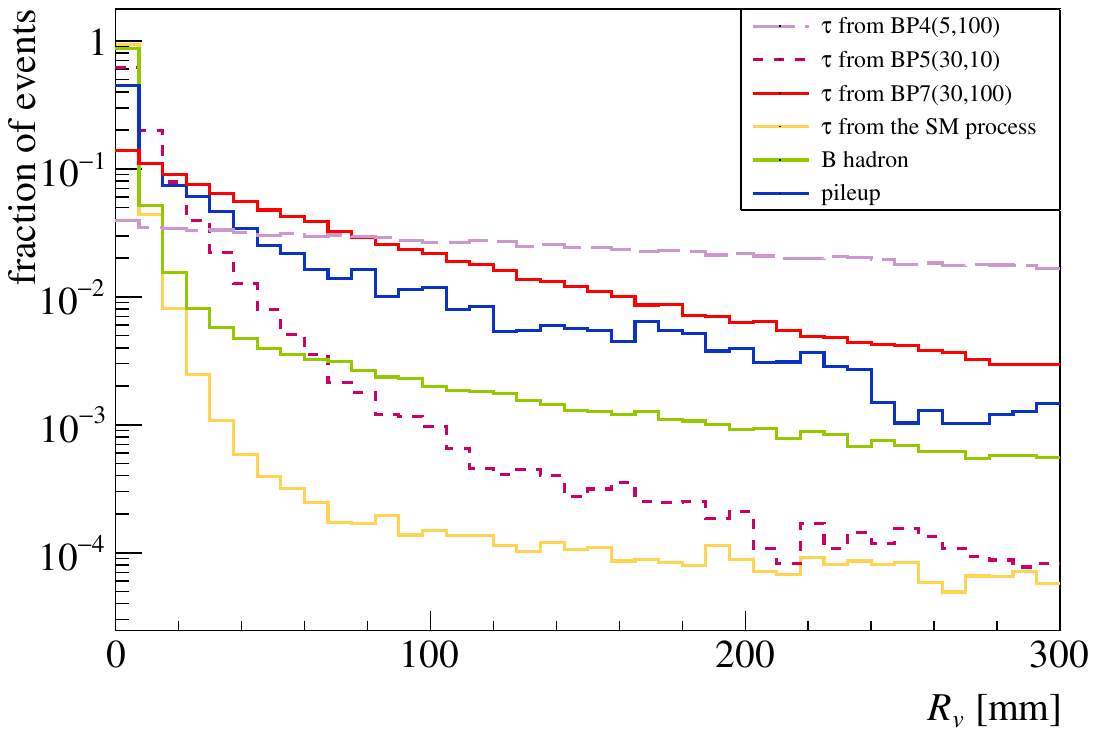}
     \vspace{-.5cm}
	\caption{The $R_v$ (radius of production position) distributions of events: yellow for prompt tau leptons from the SM process, green for $B$-hadrons, blue for pileup and other colors for tau leptons from pseudo-scalar decay.}
\label{vRprod}
\end{figure}

The gluon-gluon fusion Higgs production cross section is scaled to the N$^3$LO value~\cite{LHCHiggsCrossSectionWorkingGroup:2016ypw}.
For the $t\bar{t}$ background, we require the top quark to decay into a $W$-boson and a $b$ quark, with the $W$-boson subsequently decaying leptonically, while the $t\bar{t}$ cross section is scaled to the NNLO+NNLL value~\cite{Czakon:2011xx}.
Among the $Z\tau\tau$ backgrounds, the $Z\gamma^\ast$ component is estimated to be about $38\%$ by \textsc{MadGraph5\_aMC@NLO}. 
As a crosscheck, we show the distributions of the radius of particle production position in Fig.\ref{vRprod}, which  demonstrates that these samples are generated according to physics expect. 

\begin{table}
\begin{center}
\setlength\tabcolsep{.6em}
\renewcommand{\arraystretch}{1.3}
\caption{Physics parameters for benchmark points in the numerical study. The tan$\beta$ is taken at $0.8$ ($1.3$) for low (high) masses, $m^2_{12}$ is taken at $3900$ ($4200$) GeV$^2$ for low (high) masses.  The value of $\cos(\beta-\alpha)$ is 0.03 for all points.}
\vspace{.2cm}
\begin{tabular}{ l | c  c  c  c | c  c  c c }
\hline
 ~ & BP1 & BP2 & BP3 & BP4 & BP5 & BP6 & BP7 & BP8 \\
 \hline
$m_A$ (GeV)   & \multicolumn{4}{c|}{5} & \multicolumn{4}{c}{30}  \\ 
\hline
$c\tau$ (mm)  & 3 & 10 & 30 & 100 & 10 & 30 & 100 & 300 \\
$\kappa_\ell$ ($10^{-6}$)  & 95 & 52 & 30 & 16 & 18 & 10 & 5.6 & 3.3 \\
\hline
\end{tabular}

\label{bps} 
\end{center}
\end{table}

\subsection{The fast simulation and selection procedure}
The framework \textsc{Delphes} \cite{deFavereau:2013fsa} is adopted in this work. The setup of detector responses mainly follows that of ATLAS with pileup therein. New modules are developed as described below to accommodate a long-lived particle, in consistency with the methodology described in \cite{ATLAS:2023nze,CMS:2022prd,atlas:2019pn,ATLAS:2019wqx,CMS:2021tkn}\footnote{These algorithms and the response parameters therein are followed with the working point Medium employed as default, just for a quick estimation of physics potential.}:
\begin{itemize}
\item[(1)] {\bf Displaced tracks.}
In the scenario of LLP with a lifetime around dozens of millimeters, tracks from the LLP decay will be produced within the Tracker but somewhat far away from the collision point. Typically this can be reflected by the transverse impact parameter $d_0$ as shown in Fig.\ref{trkIP}. In our tracking module of \textsc{Delphes}, $d_0 \ge 2.5$ mm is adopted to define the displaced tracks, while other setup as in \cite{ATLAS:2023nze}. There is also an observation that the mass difference around 35 GeV $(\simeq m_h - m_Z )$ leaves only less energy to tracks, so a low threshold $p_T \ge 0.5$ GeV is used for the track transverse momentum.  But a track produced too far ($R_v \ge 300 ~mm$) will leave too few effective hits within the Inner Detector and thus will be rejected.  From Figs.\ref{vRprod} and \ref{trkIP} we see that with a fixed $c\tau$ value a lighter pseudo-scalar is more likely to travel a longer distance because it tends to have a higher velocity.     

\begin{figure}[!t]
	\centering
	\includegraphics[width=12cm]{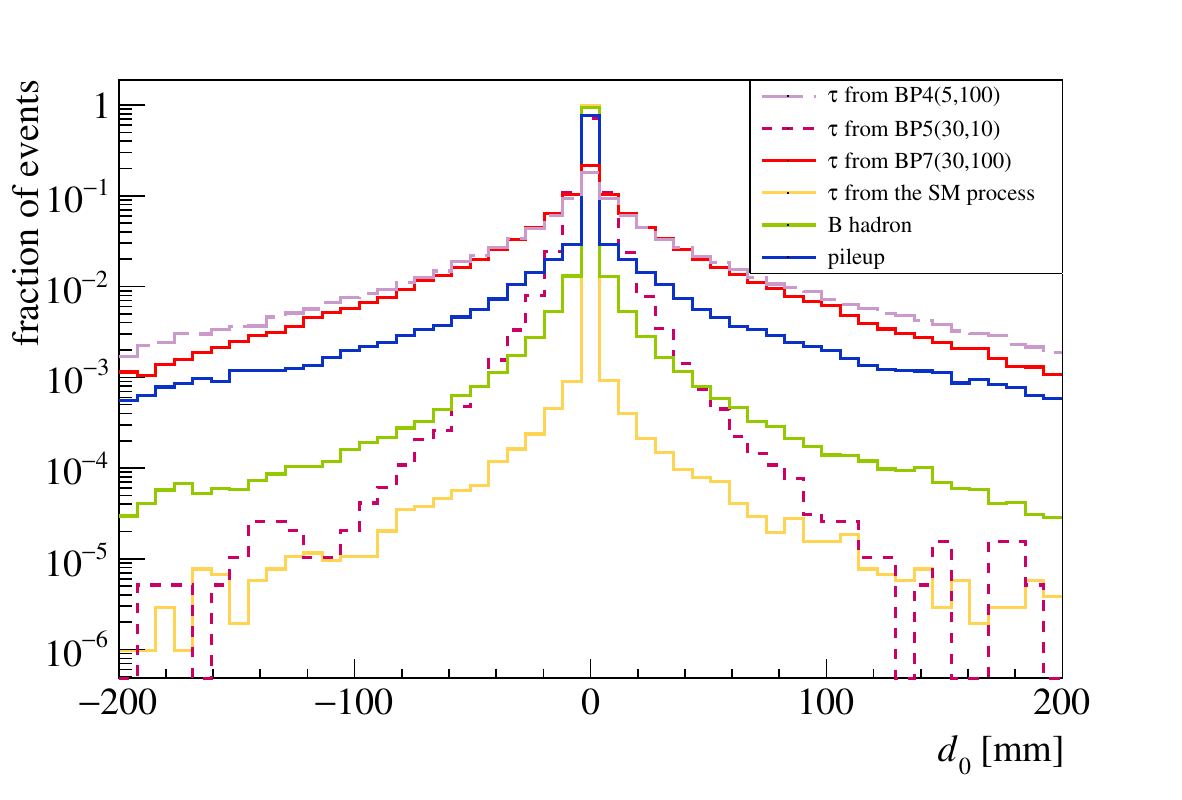}
  \vspace{-.5cm}
  \caption{The $d_0$-distributions of events: yellow for SM prompt tracks, green for $B$-hadrons,  blue for pileup and other colors denote signals. }

\label{trkIP}
\end{figure}

\item[(2)] {\bf Displaced $\tau$ jet ($\tau$ jet with displaced tracks).} 
A tau jet basing on the performance described in \cite{atlas:2019pn} is expected to be associated with one or three extra displaced tracks within a cone $\sqrt{{(\Delta\phi)}^2 + {(\Delta\eta)}^2} \le 0.2$,  as aforementioned. To keep as many tau signal events as possible, the threshold of tau $p_T$ is reduced to 10 GeV from the conventional cut like 20 GeV. For the taus with $p_T$ below 20 GeV, another (tight) working point \cite{atlas:2019pn,CMS:2022prd} is adopted for their identification, then the possibility for a b-jet to be mis-identified as a tau jet is controlled to about $10\%$. This tighter choice also brings an efficiency degradation for tau reconstruction roughly by $60\%$.  For the events with only one hadronic tau tagged, a displaced muon is required as a substitution. 

\item[(3)] {\bf Displaced joint vertex.}
Some seed vertices are reconstructed from (displaced) tracks. These candidate vertices are expected to have large distances from interacting points as shown in Fig.\ref{vRprod}. 
Furthermore, the decaying vertices from two taus are expected to be close to each other due to their low velocities, 
so that we manage to reconstruct a joint vertex by merging two taus in this work. 
In Fig.\ref{VdelZ} the averaged minimal distance between vertices along z-direction is presented, which shows that the pileup can be much suppressed when only a small slice is kept\footnote{For more dense pileup, their vertices distance will become smaller and contaminate more.}. With the cuts on the distances between vertex seeds, e.g. $( \Delta r < 4~ mm,~\Delta z < 2~ mm )$, we find that the backgrounds (mainly Pileup) can be suppressed to few percent level, even though this new attempt differs from the conventional vertexing. Also, the displaced vertex has strong correlation to the presence of displaced tau.

 \begin{figure}[!t]
	\centering
	\includegraphics[width=12cm]{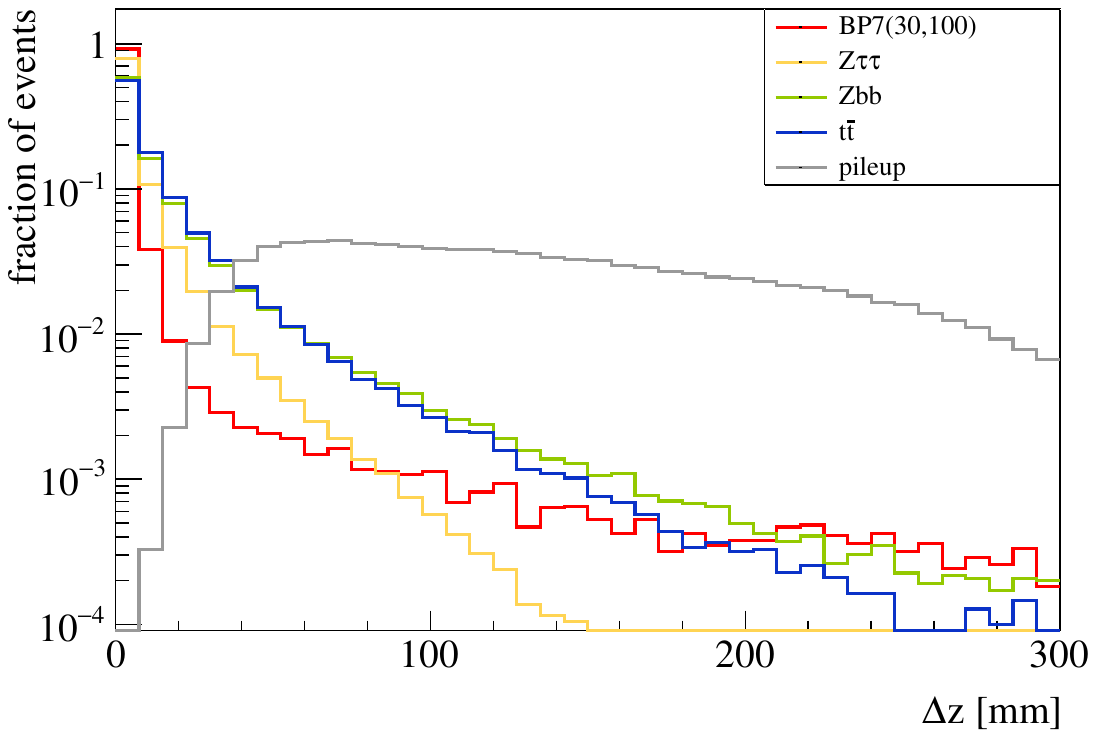}
  \vspace{-.5cm}
	\caption{The distribution of the average minimal distance between vertices in z-direction.}
\label{VdelZ}
\end{figure}

\item[(4)]  {\bf Reconstruction of $Z$-boson mass and visible mass of Higgs boson.}
As usual, the invariant mass for a pair of same-flavor and opposite-sign leptons is required to fall in a window of $81{\rm ~GeV} < m_{\ell \ell} < 101 {\rm ~GeV}$, with isolated leptons in the phase space of $p_T^\ell > 15{\rm ~GeV}$ and $|\eta| < 2.5$. No doubt this $Z$-peak (in fact the leptons' isolation) can suppress the QCD backgrounds down to a sufficiently low level. 
Also, by adding up the four-momenta of a pair of charged leptons (from $Z$) and a pair of tau candidates, a visible mass is reconstructed to suppress events other than the Higgs decay. As shown in Fig.\ref{visMh}, it helps to reject all backgrounds. 
At last, the leptons from Z provide a very good trigger in actual experiments.

 \begin{figure}[!t]
	\centering
\includegraphics[width=12cm]{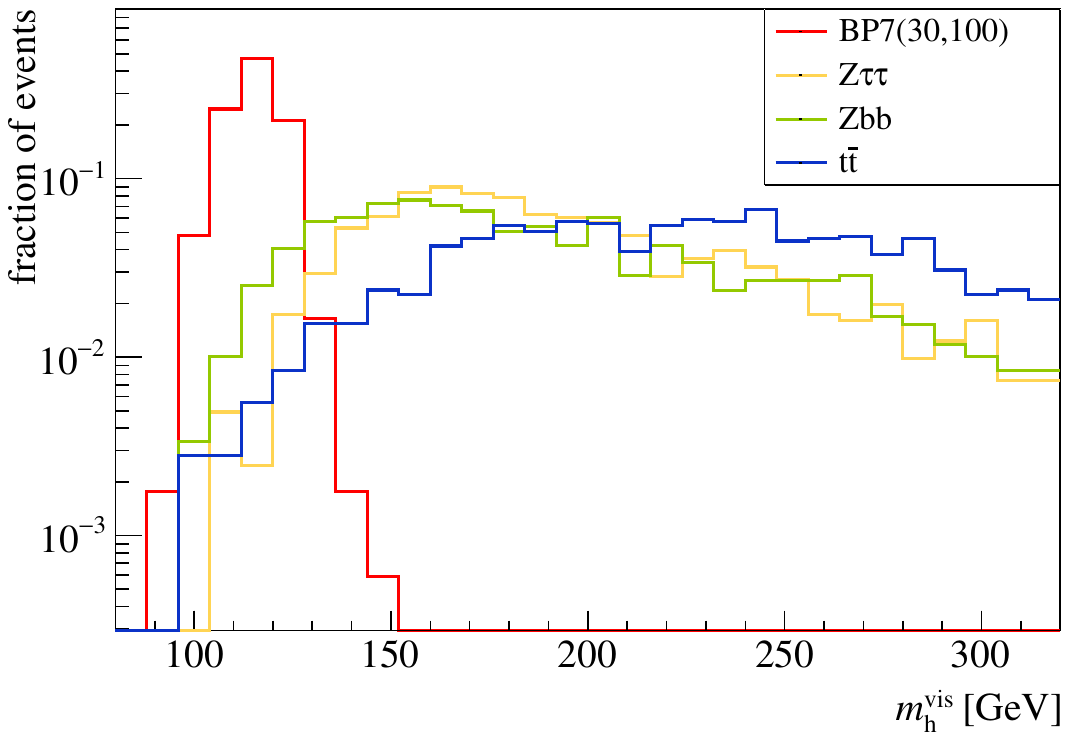}
  \vspace{-.5cm}
	\caption{The distribution of visible mass reconstructed from di-leptons associated with a pair of taus. }.
\label{visMh}
\end{figure}

\end{itemize}

\begin{table}
\begin{center}
\setlength\tabcolsep{.6em}
\renewcommand{\arraystretch}{1.3}
\caption{The separately estimated efficiencies for the backgrounds and the signals at each benchmark points. And the cross-sections prior to cuts of each backgrounds and signal (with the assumption $BR(h\to ZA)=1\%$ and $BR(A\to\tau\tau)=100\%$) are shown in the second column.} 

\small
\begin{tabular}{ l |c |c c c c c}
\hline
  ~ & \thead{cross-section \\ (pb)} & $m_Z$-window           &   \thead{displaced \\ $\tau$-jet}      &   \thead{displaced \\ vertex}   &  $m^{vis}_h \le 125$ GeV   &   \thead{final \\ efficiency} \\
 \hline 
  $Z \tau\tau$ & 0.027&  0.4                     & $1.4\times {10}^{-3}$   & $7.9\times {10}^{-2}$  & $7.5\times {10}^{-3}$    & $3.3\times {10}^{-7}$  \\
  $ Zbb$ & 68.2       & 0.4                     &  $9.8\times {10}^{-5}$  & $6.9\times {10}^{-2}$  & $9.9\times {10}^{-2}$    & $4.1\times {10}^{-7}$ \\ 
  $t{\bar t}$   & 92.4 & $2.2\times {10}^{-2}$   & $2.5\times {10}^{-4}$   & $9.1\times {10}^{-2}$   &  $2.5\times {10}^{-2}$  &$ 1.7\times {10}^{-8}$  \\ 

  \hline
  BP1  &\multirow{8}*{\thead{0.014, \\assuming\\$BR(h\to$\\$ ZA)=1\%$,\\$BR(A\to $\\$ \tau \tau)=100\%$.}} &0.44   &$3.2\times {10}^{-3}$    &$8.2\times {10}^{-2}$    & 0.95  & $0.051\%$  \\ 

  BP2  & &0.44   &$6.9\times {10}^{-3}$    &0.16    & 0.94  & $0.15\%$  \\ 
  BP3  & &0.44   &$5.6\times {10}^{-3}$    &0.14    & 0.95  & $0.11\%$  \\
  BP4  & &0.44   &$2.5\times {10}^{-3}$    &$7.8\times {10}^{-2}$    & 0.93  & $0.035\%$  \\
  \cline{1-1} \cline{3-7}
  BP5  & &0.36   &$1.3\times {10}^{-2}$    &0.19    & 0.95  & $0.26\%$  \\
  BP6  & &0.36   &$1.8\times {10}^{-2}$    &0.31    & 0.95  & $0.36\%$  \\
  BP7  & &0.36   &$1.9\times {10}^{-2}$    &0.34   & 0.97  & $0.43\%$  \\
  BP8  & &0.36   &$1.2\times {10}^{-2}$    &0.24   & 0.98  & $0.30\%$  \\
\hline
\end{tabular}

\label{SelPerf} 
\end{center}
\normalsize 
\end{table}

The performance of these simulation procedures is summarized in Table \ref{SelPerf}. For illustration, the efficiency is shown independently at each step. In fact, there are correlations among these cuts, especially between displaced tau-jet and displaced vertex, so the final efficiency differs from the simple product of these steps. 
In short, the signal events can be selected at a few thousandths, while the backgrounds are effectively (mainly by displaced tracks) reduced to about 4 or 5 orders lower. 
By a closer glance at these benchmark points, the efficiency for the high mass $(30\rm GeV)$ can reach the maximal value with $c\tau\simeq 100{\rm ~mm}$ ($0.43\%$ for BP7).
In comparison, the efficiencies for a lighter pseudo-scalar are relatively lower: it either contributes mainly softer taus, or goes beyond the sensitive zone since it travels faster (e.g. BP4).
During these selection procedures, we find that the pileup entered by contaminating displaced tracks, playing a dominant role in backgrounds. 

Our study can be applied to other models if they have the same event signatures.
Upon these signatures, a future lepton collider like CEPC or FCCee will give better performance for the absence of pileup. But this goes beyond the scope of this work.

\subsection{Results and discussions}
\label{results}

 We now investigate the simulation result with an integrated luminosity $\mathcal{L} = 300~fb^{-1}$ 
 (with pileup multiplicity around 80) at the 13 TeV LHC. 
 In addition to the statistical uncertainty (estimated at this luminosity), a total systematical error is assumed at a similar level.
 
\begin{figure}[!t]
	\centering
\includegraphics[width=11cm,height=10cm]{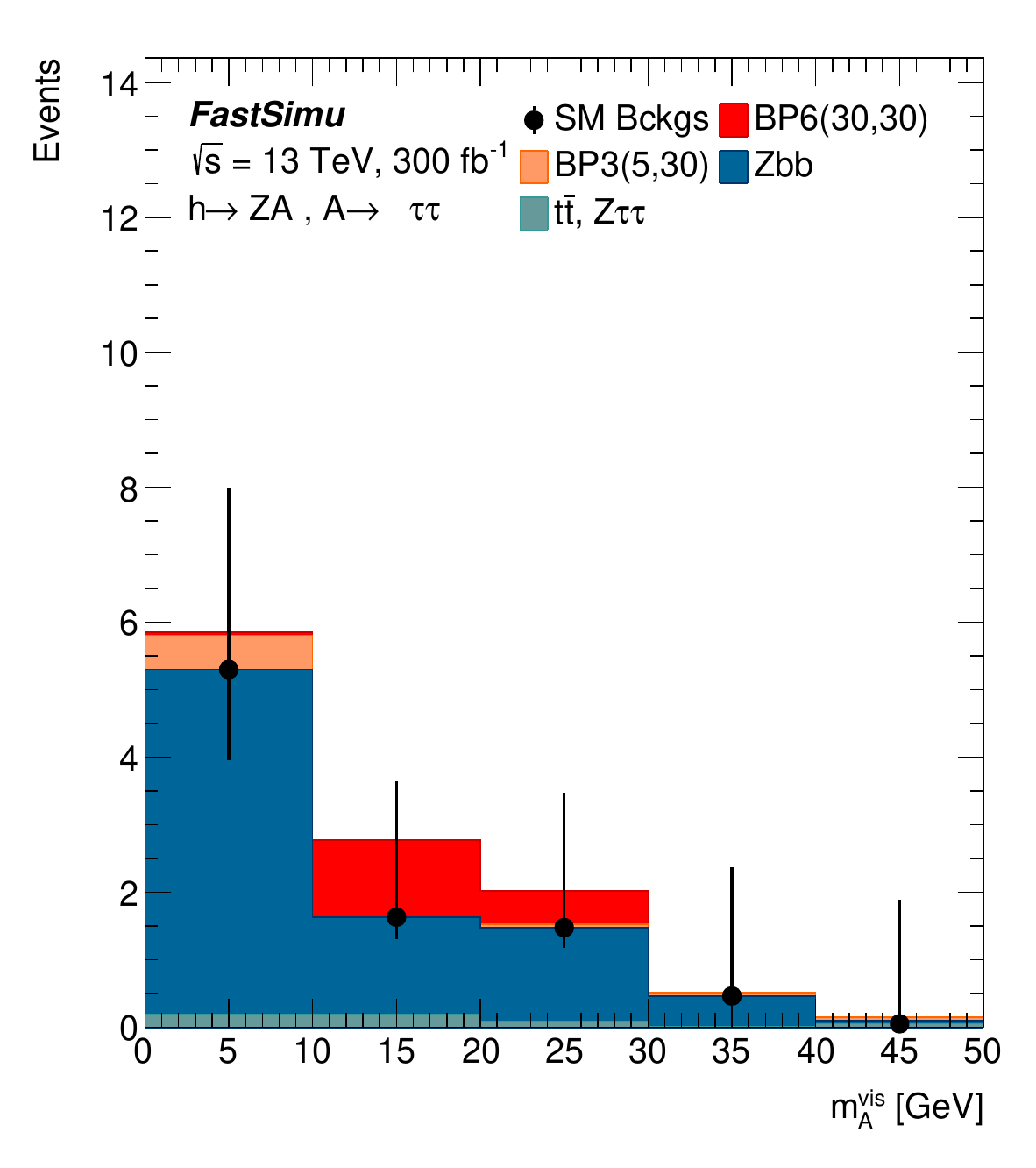}
  \vspace{-.5cm}
	\caption{ $BR(h\to ZA)$ is presumed at ${10}^{-3}$ just for a schematic illustration of signal event yields,  with the red 
 for $(m_A=30{\rm ~GeV},~c\tau=30 {\rm ~mm})$ and orange for $(m_A=5 {\rm ~GeV},~c\tau=30  {\rm ~mm})$. Other colors denote the backgrounds.}.
\label{yields}
\end{figure}

 With the procedure described in the above subsection, it is straightforward to estimate the expected yields of signal and backgrounds at the LHC, with a distribution of the (visible) invariant mass of tau candidates as in Fig.\ref{yields}. An obvious observation is that the $Z(\to l^+ l^-)+b\bar{b}$ is the main background, especially in the range of low $m_A^{vis}$, which will make the signal with low $m_A$ (orange in this figure) more difficult to detect.  

\begin{table}
\begin{center}
\setlength\tabcolsep{.6em}
\renewcommand{\arraystretch}{1.3}
\caption{Expected observation upon pure backgrounds for aforementioned benchmark points. 
The first row is the expected limit on  $BR(h\to ZA)$ estimated with pure backgrounds,
while the second row is the expected $BR(h\to ZA)$ to reach the expected $3\sigma$ significance through the selection procedures. }

\vspace{.3cm}
\begin{tabular}{ l | c  c c c | c c c c}
\hline
$( m_A, c\tau )$      &  (5,3)  &   (5,10)  & (5,30)  & (5,100) & (30,10)  &  (30,30)  & (30,100) & (30,300)  \\ 
\hline
limit           &  2.57$\%$     &  0.87$\%$      &  1.20$\%$      &  3.74$\%$       &  0.41$\%$       &    0.29$\%$    &   0.25$\%$       &   0.35$\%$    \\
at $3\sigma$   &  12$\%$      &  3.9$\%$        &  5.4$\%$       &  17$\%$    &  2.3$\%$        &   1.6$\%$   &  1.3$\%$   & 2.0$\%$   \\
 \hline
\end{tabular}
\label{tab3} 
\end{center}
\end{table}

The branching ratio $BR(h\to ZA)$ can be regarded as a free parameter in this work.
In the case that no excess will be observed but only these backgrounds, an expected limit for the branching ratio can be estimated with the $CL_s$ ($95\%$) procedure  \cite{statPractice:2015,TrexFitter:2020}, as shown in Table \ref{tab3}. In the same condition, when the branching ratio is sufficiently large (as listed in this table), the signal significance can reach the $3\sigma$ level\footnote{With a rough estimation by $2(\sqrt{B+S}-\sqrt{B})\frac{B}{B+ {\Delta B}^2}$}.  

\section{Conclusions}
\label{conclusion}
We studied the Higgs decay into a $Z$-boson plus a long-lived light pseudo-scalar which dominantly decays to tau leptons after flying over a distance from the production point and presents a displaced vertex in an Inner Detector of a generally purposed experiment like ATLAS or CMS. 
We focused on the LHC experiment and performed \textsc{Delphes} simulations for the signal and backgrounds. For the signal we  took the aligned two-Higgs-doublet model as an example and  demonstrated some benchmark points. We found that with the luminosity of 300  fb$^{-1}$ the signal is detectable when the branching ratio of $h\to ZA$ reaches a few percents. So our study suggests an experimental search for such a process in the ongoing LHC. 

\addcontentsline{toc}{section}{Acknowledgments}
\acknowledgments
This work was supported by the National Natural Science Foundation of China (12135014, 11821505, 12075300, 12335005, 11975013), by the Research Fund for Outstanding Talents from Henan Normal University (5101029470335), by the Peng-Huan-Wu Theoretical Physics Innovation Center (12047503) funded by the National Natural Science Foundation of China and by the project ZR2023MA038 supported by Shandong Provincial Natural Science Foundation. 

\appendix
\section{Some relations in the parameter space of 2HDM}
\label{parameter-space}

It is convenient to take the masses as inputs and express the couplings in terms of them:
\begin{eqnarray}\label{poten-cba}
 &&v^2 \lambda_1  = \frac{m_H^2 c_\alpha^2 + m_h^2 s_\alpha^2 - m_{12}^2 \tan\beta}{ c_\beta^2}, \ \ \ 
v^2 \lambda_2 = \frac{m_H^2 s_\alpha^2 + m_h^2 c_\alpha^2 - m_{12}^2/ \tan\beta}{s_\beta^2},  \nonumber \\  
&&v^2 \lambda_3 =  \frac{(m_H^2-m_h^2) s_\alpha c_\alpha + 2 m_{H^{\pm}}^2 s_\beta c_\beta - m_{12}^2}{ s_\beta c_\beta }, \ \ \ 
v^2 \lambda_4 = \frac{(m_A^2-2m_{H^{\pm}}^2) s_\beta c_\beta + m_{12}^2}{ s_\beta c_\beta },  \nonumber \\
 &&v^2 \lambda_5=  \frac{ - m_A^2 s_\beta c_\beta  + m_{12}^2}{ s_\beta c_\beta } \, . 
 \label{eq:lambdas}
\end{eqnarray}
The parameters $m_{11}^2$ and $m_{22}^2$ are determined by the potential minimization conditions,
\bea
&&\quad m_{11}^2 = m_{12}^2 \tan\beta - \frac{1}{2} v^2 \left( \lambda_1 c_\beta^2 + (\lambda_3+\lambda_4+\lambda_5) s_\beta^2 \right)\,,\\
&& \quad m_{22}^2 =  m_{12}^2  \cot\beta - \frac{1}{2} v^2 \left( \lambda_2 s_\beta^2 + (\lambda_3+\lambda_4+\lambda_5) c_\beta^2 \right)\, .
\label{min_cond}
\eea

\section{Alignment in flavour space}
\label{aligned}

The alignments in flavour space are explicitly recast as,
 \bea\label{align-kf}
 &&(X_{u1})_{ii}=\frac{\sqrt{2}m_{ui}}{v}(c_\beta-s_\beta \kappa_u),~~~~~(X_{u2})_{ii}=\frac{\sqrt{2}m_{ui}}{v}(s_\beta+c_\beta \kappa_u),\nonumber\\
&&(Y_{\ell 1})_{ii}=\frac{\sqrt{2}m_{\ell i}}{v}(c_\beta-s_\beta \kappa_\ell),~~~~~(Y_{\ell 2})_{ii}=\frac{\sqrt{2}m_{\ell i}}{v}(s_\beta+c_\beta \kappa_\ell),\nonumber\\
&&(X_{d1})_{ii}=\frac{\sqrt{2}m_{di}}{v}(c_\beta-s_\beta \kappa_d),~~~~~(X_{d2})_{ii}=\frac{\sqrt{2}m_{di}}{v}(s_\beta+c_\beta \kappa_d),
\eea
with
\beq 
X_{u1,2}=V_{uL}^\dagger Y_{u1,2} V_{uR},~~
X_{d1,2}=V_{dL}^\dagger Y_{d1,2} V_{dR},~~
V_{CKM}=V_{uL}^\dagger V_{dL}.
\eeq
where $i=1,2,3$ is the index of generation and all the off-diagonal elements are zero. 
$V_{uL,R}$ ($V_{dL,R}$) are unitary matrices which transform the interaction
eigenstates to the mass eigenstates for the left-handed and right-handed up-type (down-type) quark fields. From Eqs. (\ref{rot-cp-odd}), (\ref{gene-Yukawa}) and (\ref{align-kf}), we can obtain the Yukawa couplings.

\section{Constraints on $\tan\beta$  }
\label{tbeta}

When $\cos(\beta-\alpha)$ is very close to 0, the coupling constants in Eq. (\ref{poten-cba}) can be approximated by
\begin{eqnarray}
v^2 \lambda_1 &=&  m_h^2 - \frac{t_\beta^3\,(m_{12}^2 -m_H^2  s_\beta c_\beta ) }{ s_\beta^2}\,, \\
v^2 \lambda_2 &=& m_h^2 - \frac{ (m_{12}^2 -m_H^2  s_\beta c_\beta) }{ t_\beta s_\beta^2 }\,,\\
\label{eq:alignment2_lambda}
v^2 \lambda_3 &=&  m_h^2 + 2 m_{H^{\pm}}^2 - 2m_H^2 -  \frac{t_\beta (m_{12}^2 -m_H^2  s_\beta c_\beta)}{  s_\beta^2}\,,\\
v^2 \lambda_4 &=&  m_A^2-  2 m_{H^{\pm}}^2 + m_H^2+  \frac{t_\beta (m_{12}^2 -m_H^2  s_\beta c_\beta)}{  s_\beta^2}\,,\\
v^2 \lambda_5 &=&  m_H^2 - m_A^2+  \frac{t_\beta (m_{12}^2 -m_H^2  s_\beta c_\beta)}{  s_\beta^2} \,.
\label{poten-cba0}
\end{eqnarray}
The vacuum stability demands the following conditions \cite{Deshpande:1977rw}
\beq\label{vacuum-condi}
\lambda_1 > 0,~~\lambda_2 > 0,~~\lambda_3 + \sqrt{\lambda_1\lambda_2} > 0,~~\lambda_3 + \lambda_4 - \mid\lambda_5\mid +\sqrt{\lambda_1\lambda_2} > 0.
\eeq
If $\tan\beta$ is large, the perturbativity requirement of $\lambda_1$ favors $m^2_{12}- m^2_H s_\beta c_\beta\to 0$. However, when both $\cos(\beta-\alpha)=0$ and $m^2_{12}- m^2_H s_\beta c_\beta =$ 0 hold exactly, according to Eq.~(\ref{poten-cba0}), the last condition in Eq. (\ref{vacuum-condi}) leads to a stringent bound 
\beq\label{mass-hAh2}
m_h^2+m_A^2> m_H^2,
\eeq
which is contradictory to a light $A$. Therefore, in the scenario of a light $A$ and $\cos(\beta-\alpha)=0$, $\tan\beta$ is disfavored to be large unless the parameters are  fine-tuned .    

\addcontentsline{toc}{section}{References}
\bibliographystyle{JHEP}
\bibliography{ref}

\end{document}